\documentclass[preprint,prd,nofootinbib]{revtex4}
\usepackage{amsmath}
\usepackage{amssymb}
\usepackage{graphicx}
\usepackage{subfigure}
\newcommand{\nc}{\newcommand}
\nc{\beq}{\begin{equation}}
\nc{\eeq}{\end{equation}}
\nc{\beqa}{\begin{eqnarray}}
\nc{\eeqa}{\end{eqnarray}}

\long\def\sl#1{\hbox{\tiny \it #1}}			
\long\def\teq#1{\hbox{$#1$}}				

\nc{\vv}{\boldsymbol}					
\nc\La{{\cal L}}
\nc\Ha{{\cal H}}
\nc{\bert}{\raise-0.45mm\hbox{\Large$\Box$}}	
\nc\A{A^\mu}
\nc{\PgP}{\bar\psi\gamma^\mu\psi}			
\nc{\tensorbilinear}{\bar\psi\frac{i}{2}(\gamma_\mu
\buildrel\rightarrow\over\partial_\nu -
\gamma_\mu\buildrel\leftarrow\over\partial_\nu)\psi}

\def\gsim{\mathrel{\rlap{\lower4pt\hbox{$\sim$}}
    \raise1pt\hbox{$>$}}}					
\def\lsim{\mathrel{\rlap{\lower4pt\hbox{$\sim$}}
    \raise1pt\hbox{$<$}}}					

\def\Dsl{\,\raise.15ex \hbox{/}\mkern-12.8mu D}
\newcommand\Tr{{\rm Tr\,}}

 \def\lsim{\mathrel{\rlap{\lower4pt\hbox{\hskip1pt$\sim$}}
\raise1pt\hbox{$<$}}}

\long\def\ata#1{\footnote{E-mail: {\tt #1}}}
\long\def\adr#1{$^{\rm #1}$}

\def\gsim{\mathrel{\rlap{\lower4pt\hbox{\small \hskip1pt$\sim$}}
 \raise1pt\hbox{$>$}}}

\def\lsim{\mathrel{\rlap{\lower4pt\hbox{\small \hskip1pt$\sim$}}
 \raise1pt\hbox{$<$}}}
 
\newwrite\ffile\global\newcount\figno \global\figno=1

\def\writedef#1{}
\def\figin{\epsfcheck\figin}\def\figins{\epsfcheck\figins}

\def\epsfcheck{\ifx\epsfbox\UnDeFiNeD \message{(NO epsf.tex, FIGURES
WILL BE IGNORED)}
\gdef\figin##1{\vskip2in}\gdef\figins##1{\hskip.5in}
instead \else\message{(FIGURES WILL BE INCLUDED)}%
\gdef\figin##1{##1}\gdef\figins##1{##1}\fi} \def\figinsert{}
\def\ifig#1#2#3{\xdef#1{fig.~\the\figno} \writedef{#1\leftbracket
fig.\noexpand~\the\figno}%
\figinsert\figin{\centerline{#3}}\medskip\centerline{\vbox{\baselineskip12pt
\advance\hsize by -1truein\center\footnotesize{ Fig.~\the\figno.} #2}}
\bigskip\endinsert\global\advance\figno by1}
\def\endinsert{}

\begin{document}

\title{~ \\ The Story of $\cal O$:\\ Positivity constraints in effective field theories}

\author{Alejandro~Jenkins\adr{a,}\adr{b}\ata{ajv@mit.edu} and Donal~O'Connell\adr{a}\ata{donal@theory.caltech.edu}}

\affiliation{\adr{a}Department of Physics \\ California Institute of Technology, Pasadena, CA 91125}

\affiliation{\adr{b}Center for Theoretical Physics \\ Laboratory for Nuclear Science \\ Massachusetts Institute of Technology \\ Cambridge, MA 02139 \\ ~}

\preprint{CALT-68-2607}
\preprint{MIT-CTP-3764}

\begin{abstract}

\bigskip

We propose a simple method for identifying operators in effective field theories whose coefficients must be positive by causality.  We also attempt to clarify the relationship between diverse positivity arguments that have appeared in the literature.  We conjecture that the superluminal perturbations identified in non-positive effective theories are generally connected to instabilities that develop near the cutoff scale.  We discuss implications for the ghost condensate, the chiral Lagrangian, and the Goldstone bosons of theories with spontaneous Lorentz violation.

\end{abstract}

\maketitle


\newpage

\section{Introduction}

A variety of classical and quantum arguments have been formulated to require the positivity of the coefficients of higher-dimensional (i.e., irrelevant) operators in effective field theories, including General Relativity (see, for instance, \cite{barrow, mullerschmidt, ANSAR, trodden}).  Here we will make some remarks which help to clarify which operators one expects to obey positivity constraints, as well as the connection between the diverse positivity arguments in the literature.  In particular, we will argue that positivity is expected to follow from causality for operators of the form
\beq
{\cal O} \propto
{\cal O}_1^{\mu_1 \ldots \mu_j}
{\cal P}^{(j)}_{\mu_1 \ldots \mu_j; \nu_1 \ldots \nu_j}
{\cal O}_1^{\nu_1 \ldots \nu_j}~,
\eeq
where \teq{{\cal O}_1} contains sufficient derivatives and \teq{{\cal P}^{(j)}_{\mu_1 \ldots \mu_j; \nu_1 \ldots \mu_j}} is the zero-momentum propagator for a massive, spin-$j$ mediator.  We do not expect positivity constraints from causality alone for operators of the form
\beq
{\cal O} \propto
{\cal O}_1^{\mu_1 \ldots \mu_j}
{\cal P}^{(j)}_{\mu_1 \ldots \mu_j; \nu_1 \ldots \nu_j}
{\cal O}_2^{\nu_1 \ldots \nu_j}~.
\eeq

Furthermore, we shall argue that theories with such \teq{{\cal O}}'s that violate positivity do not admit stable, perturbative UV completions, and that the instabilities near the cutoff scale of non-positive effective theories are associated to the superluminal modes that may appear in the IR.  We shall comment on the implications of this for the ghost condensate mechanism that has been proposed as a model of gravity in a Higgs phase \cite{ghost}, for the chiral Lagrangian, and for theories in which Lorentz invariance is spontaneously broken by a VEV for a vector quantity.  This discussion is motivated principally by \cite{ANSAR}, whose notation we will adopt.

\section{Superluminality and analyticity}

Consider the Lagrangian
\beq
\La = \frac 1 2 \left( \partial_\mu \pi \right)^2 - \frac 1 2 m^2 \pi^2 + \frac{c_3}{2 \Lambda^4} \left( \partial_\mu \pi \right)^4 + \ldots ~,
\label{piaction}
\eeq
which could describe an effective theory, at energy scales well below $\Lambda$, for a scalar field $\pi$ with a very small mass $m$.\footnote{For $m \neq 0$ Eq. (\ref{piaction}) has no shift symmetry in $\pi$ and we would expect other self-interactions such as \teq{\pi^4}, \teq{\pi \left( \partial_\mu \pi \right)^2}, etc.  However, we will be concerned here mostly with the limit $m \to 0$.}  In \cite{ANSAR}, the authors offer two distinct causality arguments to constrain $c_3$.  The first is classical and applies for $m=0$:  Consider a background $\pi_0$ such that \teq{\partial_\mu \pi_0 = C_\mu}, for constant $C$.  For \teq{\left| C^2 \right| \ll \Lambda^4}, we obtain the linear dispersion relation
\beq
k^2 + \frac{4 c_3}{\Lambda^4} (C \cdot k)^2 = 0~,
\label{phidispersion}
\eeq
where $k$ is the 4-momentum of a plane wave of the perturbation \teq{\varphi \equiv \pi - \pi_0}, which is the non-relativistic Goldstone of the spontaneous breaking of the shift symmetry  \teq{\pi \to \pi + c} by the background.  Absence of superluminal excitations then requires $c_3 \geq 0$.  For $c_3 < 0$, the superluminal excitations are not tachyons and the background $\pi_0$ is stable, even though the Hamiltonian is not minimized by it, because shift-symmetry implies the conservation of 
\beq
Q = \int d^3 x \, \dot \pi \left[ 1 + \frac{2 c_3}{\Lambda^4} \left(\dot \pi^2 - \left| \vv \nabla \pi \right|^2 \right) \right]
\label{charge}
\eeq
and small perturbations $\varphi$ that conserve $Q$ cannot lower the energy \cite{grad, holes, holes2}.  If $c_3$ were negative, it would be possible to use these superluminal excitations to construct closed timelike curves in certain non-trivial backgrounds \cite{ANSAR}.

Let us now consider the case \teq{m \neq 0} in Eq. (\ref{piaction}).  The shift symmetry is then explicitly broken at a scale $m$, which should also be the scale of the mass of the pseudo-Goldstone $\varphi$.  Subluminality of $\varphi$ at long wavelengths is assured as long as \teq{m^2 > 0}.  We expect that absence of superluminal $\varphi$ near the cutoff scale will impose, at best, only a limit of the form
\beq
c_3 \gsim - \frac{m^2}{\Lambda^2}~.
\label{classicalm}
\eeq

The second argument in \cite{ANSAR} is based on the analyticity of the S-matrix for Eq. (\ref{piaction}).  Let ${\cal M}(s,t)$ be the amplitude for \teq{\pi \pi \to \pi \pi} scattering and consider the analytic continuation of ${\cal A} (s) \equiv {\cal M}(s, t=0)$ onto the complex plane.   For an intermediate scale $M$ such that $m \ll M \ll \Lambda$, analyticity requires that \teq{{\cal A}''(M^2)} be strictly positive.  Since \teq{{\cal A}''(M^2)} is equal to \teq{2 c_3 / \Lambda^4} plus loop corrections suppressed by \teq{M^4/\Lambda^8}, we expect a limit of the form 
\beq
c_3 > 0
\label{Smatrix}
\eeq
regardless of the value of the small mass $m$.\footnote{In fact, the S-matrix analyticity argument in \cite{ANSAR} requires the introduction of a regulator mass $m$, which may be taken to zero at the end.}  It therefore seems that the two positivity arguments in \cite{ANSAR} are not equivalent and that analyticity of the S-matrix imposes a more stringent constraint on $c_3$.

There is another important difference between the positivity argument based on Eq. (\ref{phidispersion}) and the argument from analyticity of the S-matrix.  The former identifies a violation of causality which is present already in the IR.  The latter requires closing the contour on the complex plane out at \teq{|s| \to \infty} and therefore should be interpreted as an obstruction to finding a causal UV-completion of the effective theory.  This is the spirit in which the argument has been proposed in \cite{distler} as providing a falsifiable prediction of string theory.

\section{The ghost condensate}

The positivity constraints of \cite{ANSAR} present an obstruction to the ghost condensate of \cite{ghost}.  For \teq{X \equiv \left( \partial_\mu \pi \right)^2}, the ghost condensate has an action of the form
\beq
\La = P(X)~,
\label{ghostaction}
\eeq
where $P$ is a polynomial with \teq{P'(X_\ast) = 0} and \teq{P''(X_\ast) \neq 0} at \teq{X_\ast \neq 0}.  For such an action there will generally be some background \teq{X_0 = \left(\partial_\mu \pi_0 \right)^2 = C^2} on one side of $X_\ast$ where the Goldstone $\varphi$ is superluminal.  This can be seen from the formula for the speed $v$ of linear waves in $\varphi$.  If \teq{X_0>0} then, in the frame where \teq{C^\mu = (C,0,0,0)}, we have
\beq
v^2 =   \frac{1}{1 + 2 \left| X_0 \right| P''(X_0)/P'(X_0)}
\label{speed}
\eeq
for \teq{P'(X_0) \neq 0}.\footnote{When \teq{C^\mu} is spacelike, the speed of perturbations moving along the direction of \teq{C^\mu} (in the frame where \teq{C^0 = 0}) is given by \teq{v^2 = 1 - 2 \left| X_0 \right| P''(X_0)/P'(X_0)}.  This reproduces Eq. (\ref{speed}) for \teq{\left| X_0 \right| P''(X_0)/P'(X_0) \ll 1}.}  For $X_0$ in one half-neighborhood of $X_\ast$, the quantity \teq{\left| X_0 \right| P''(X_0)/P'(X_0)} is very large and negative.  In that case $v^2 < 0$ in Eq. (\ref{speed}), signaling an instability.  But there will generally be a region where \teq{\left| X_0 \right| P''(X_0)/P'(X_0)} is small and negative, leading to \teq{v^2 > 1} and indicating the presence of stable superluminal perturbations which could be used to build closed timelike curves.

Note also that in the limit \teq{M_{\sl{Pl}} \to \infty} where the ghost condensate decouples from gravity, the overall coefficient of the action in Eq. (\ref{ghostaction}) is irrelevant.  If we normalize it to have a normal leading kinetic term \teq{X/2} then analyticity of the S-matrix for \teq{\pi \pi \to \pi \pi} forbids negative coefficients for higher powers of $X$, thus preventing the polynomial $P(X)$ from having a point \teq{P'(X_\ast) = 0} for \teq{X_\ast \neq 0}.

\section{The Story of $\cal O$}

The theory in Eq. (\ref{piaction}) with only a $c_3$ self-interaction is equivalent to
\beq
\La = \frac 1 2 \left( \partial_\mu \pi \right)^2 - \frac 1 2 m^2 \pi^2 - \frac{c_3}{2} \Lambda^2 \Phi^2 - \epsilon \frac{c_3}{\Lambda} \Phi \left( \partial_\mu \pi \right)^2~,
\label{Phiaction}
\eeq
where $\epsilon = \pm 1$, since integrating out the auxiliary field $\Phi$ corresponds exactly to substituting its equation of motion \teq{\Phi = - \epsilon \left(\partial_\mu \pi \right)^2 / \Lambda^3}.  We could therefore think of Eq. (\ref{piaction}) as describing the low-energy behavior of
\beq
\La' = \left( \frac 1 2 - \epsilon \frac{c_3}{\Lambda} \Phi \right) \left( \partial_\mu \pi \right)^2 - \frac 1 2 m^2 \pi^2 + \frac 1 2 \left(\partial_\mu \Phi \right)^2 - \frac{c_3}{2} \Lambda^2 \Phi^2 ~.
\label{Phiprop}
\eeq
Regardless of whether $m$ vanishes or not, the wrong sign of $c_3$ in Eq. (\ref{Phiprop}) leads to an instability at energy scales near the cutoff for Eq. (\ref{piaction}).  If, for the wrong sign of the $\Phi^2$ term, we attempt to make Eq. (\ref{Phiaction}) stable by adding higher-order potential terms for $\Phi$, then the corresponding low-energy effective action will still have $c_3 >0$ if $\Phi$ sits at a stable point of its potential.\footnote{It will also have operators \teq{(\partial_\mu \pi)^{2n}} for \teq{n > 1}, whose coefficients are also positive when $\Phi$ sits at a stable point.}  Transition from $c_3 > 0$ to $c_3 < 0$ in Eq. (\ref{piaction}) corresponds to the destabilization of the fixed point at which the heavy field $\Phi$ in Eq. (\ref{Phiprop}) sits.

The theory described by Eq. (\ref{Phiprop}) is not a true high-energy completion of Eq. (\ref{piaction}) because it is not renormalizable.  However, it is very simple to check that for $c_3 >0$ the forward scattering amplitudes for \teq{\pi \pi \to \pi \pi}, \teq{\pi \Phi \to \pi \Phi}, and \teq{\Phi \Phi \to \Phi \Phi} admit a perturbative UV-completion with an analytic S-matrix, since the ${\cal A}''(s)$ for all three kinds of scattering have the right sign at low energies, as required by \cite{ANSAR}.\footnote{The sign constrained by the analyticity argument in \cite{ANSAR} has become the sign of the mass-squared for the auxiliary field $\Phi$.  That is, the analyticity constraint has become a simple stability constraint.}  Then Eq. (\ref{Phiprop}) could be a good approximation, at intermediate energies, to a (perhaps fine-tuned) analytic UV-completion.\footnote{We could fine-tune \teq{|c_3| \ll 1} in Eq. (\ref{Phiprop}) so that radiative corrections leading to other couplings are under control.  Alternatively, we could control radiative corrections by taking $m$ to be small, so that the $\pi$ field has an approximate shift symmetry.}

Consider now more generally
\beq
\La' = a \Phi \cdot {\cal O}_1 - \frac{b}{2} \Lambda^2 \Phi^2~,
\label{O1}
\eeq
where ${\cal O}_1$ is some arbitrary operator and the coefficient $a$ has the appropriate mass dimension.  Then \teq{\La'} is equivalent to
\beq
\frac{a^2}{2 b}{\cal O}_1^2~.
\label{O1eq}
\eeq
The condition that $\Phi$ be non-tachyonic if it is made dynamical then imposes a constraint on the sign of \teq{{\cal O} = {\cal O}_1^2}.  This method trivially succeeds in identifying many other positivity constraints worked out in the literature, such as the positivity of $c_1$ and $c_2$ for the $U(1)$ gauge field action discussed in \cite {ANSAR}
\beq
\La = - \frac 1 4 F_{\mu \nu} F^{\mu \nu} + \frac{c_1}{\Lambda^4} \left( F_{\mu \nu} F^{\mu \nu} \right)^2 + \frac{c_2}{\Lambda^4} \left( F_{\mu \nu} \tilde F^{\mu \nu} \right)^2 + \ldots
\eeq
or the positivity of the higher-dimensional operators in General Relativity (e.g., $R^2$) discussed in \cite{barrow,mullerschmidt,trodden}.

That we expect positivity constraints on operators of the form  \teq{{\cal O} = {\cal O}_1^2} where \teq{{\cal O}_1} has enough derivatives is clear from the analyticity argument in \cite{ANSAR}, which constrains the signs of the coefficients of $s^{2n}$ in the power expansion of the forward scattering amplitude ${\cal A}(s)$, for $n \geq 1$.  We do not expect our auxiliary field method to yield a constraint on an operator without derivatives, such as \teq{-\lambda \pi^4}.  In that case
\beqa
\La' &=& - \lambda \left( 2 \Lambda \Phi \pi^2 + \Lambda^2 \Phi^2 \right) \nonumber \\
&=& -\lambda \Lambda^2 \left(\Phi + \frac{\pi^2}{\Lambda} \right)^2 + \lambda \pi^4~,
\eeqa
which is always an unstable potential, regardless of the sign of $\lambda$.  Furthermore, we should not constrain the sign of an ordinary kinetic term \teq{\kappa \left( \partial_\mu \pi \right)^2/2}.  In that case we could write
\beq
\La' = \kappa \Lambda A^\mu \left( \partial_\mu \pi \right) + \frac{\kappa}{2} \Lambda^2 A^2
\label{kinetic}
\eeq
but the coupling of the auxiliary field can be removed upon integrating by parts, since \teq{\partial_\mu A^\mu = 0} for a massive vector field.

We never expect positivity constraints from causality alone for operators of the form \teq{{\cal O} = {\cal O}_1 \cdot {\cal O}_2}.  In that case it is always possible to write
\beq
\La' = a_1 \Phi_1 \cdot {\cal O}_1 + a_2 \Phi_2 \cdot {\cal O}_2 - \frac{b_1}{2} \Lambda^2 \Phi_1^2 - \frac{b_2}{2} \Lambda^2 \Phi_2^2~,
\label{O1}
\eeq
and the sign of \teq{{\cal O}} in the equivalent action will depend on the sign of $a_1 \cdot a_2$, which we can set freely.

The method we have used to identify operators with positivity constraints amounts to a very simple-minded inverse Operator Product Expansion.  We take an operator $\cal O$ and split it up into two parts joined by a massive, zero-momentum mediator. We expect a positivity constraint if the theory that results after the mediator is made dynamical is stable if and only if the coefficient of $\cal O$ was positive.  We do not expect a constraint from causality if the theory can be made stable regardless of the sign of $\cal O$.

\section{The chiral Lagrangian}

Let us now consider how our method applies to the coefficients of the next-to-leading order operators of the $SU(2)$ chiral Lagrangian
\beqa
\La &=& \frac 1 4 v^2 \Tr \left( \partial^\mu \Sigma^\dagger \partial_\mu \Sigma \right) +
\frac 1 4 m^2 v^2 \Tr \left( \Sigma^\dagger + \Sigma \right) \nonumber \\
&& + \frac 1 4 \ell_1 \left[ \Tr \left(  \partial^\mu \Sigma^\dagger \partial_\mu \Sigma \right) \right]^2
+ \frac 1 4 \ell_2 \left[ \Tr \left( \partial^\mu \Sigma^\dagger \partial^\nu \Sigma \right) \right]
\left[  \Tr \left( \partial_\mu \Sigma^\dagger \partial_\nu \Sigma \right) \right] + \ldots~
\label{chiral}
\eeqa
where \teq{\Sigma (x) \equiv \exp \left( i \pi^i(x) \sigma^i / v \right)}, with $\pi^i$ being the pion fields and $\sigma^i$ the Pauli matrices.  We shall see that this example illustrates a subtlety which may appear when considering operators of the form \teq{{\cal O} = {\cal O}_1^{\mu_1 \ldots \mu_j}{\cal O}_{1~\mu_1 \ldots \mu_j}} for \teq{j >1}.

The Lagrangian in Eq. (\ref{chiral}) can be rewritten as
\beqa
\La &=& \frac 1 4 v^2 \Tr \left( \partial^\mu \Sigma^\dagger \partial_\mu \Sigma \right) +
\frac 1 4 m^2 v^2 \Tr \left( \Sigma^\dagger + \Sigma \right) \nonumber \\
&& + \frac{1}{4} \left[ \ell_2 P^{(2)}_{\mu \nu; \rho \sigma} + \left( \frac{\ell_2}{D-2}
+ \ell_1 \right) P^{(0)}_{\mu \nu; \rho \sigma} \right] \left[ \Tr \left( \partial^\mu \Sigma^\dagger \partial^\nu \Sigma \right) \right]
\left[  \Tr \left( \partial^\rho \Sigma^\dagger \partial^\sigma \Sigma \right) \right] + \ldots
\eeqa
where
\beq
P^{(2)}_{\mu \nu; \rho \sigma} \equiv \frac 1 2 \left( g_{\mu \rho} g_{\nu \sigma} + g_{\mu \sigma} g_{\nu \rho} - \frac{2}{D-2} g_{\mu \nu} g_{\rho \sigma} \right)
\eeq
gives the index structure, in $D$ space-time dimensions, of the propagator for a massive spin-2 particle with zero momentum, while
\beq
P^{(0)}_{\mu \nu; \rho \sigma} \equiv g_{\mu \nu} g_{\rho \sigma}~.
\eeq
We can therefore write an action with spin-2 and spin-0 auxiliary fields which reproduces Eq. (\ref{chiral}).  Making those fields non-tachyonic requires
\beq
\left\{ \begin{array} {l} \ell_2 > 0 \\ \ell_1 + \frac{\ell_2}{D-2} > 0 \end{array} \right.~.
\eeq
For $D \geq 3$, this implies that
\beq
\left\{ \begin{array} {l} \ell_2 > 0 \\ \ell_1 + \ell_2 > 0 \end{array} \right.~,
\eeq
which agrees with the constraints obtained in \cite{distler} from avoiding superluminal perturbations around the classical background \teq{\Sigma = \exp \left(i c \cdot x \sigma^3 \right)} in the limit that \teq{m \to 0}.

\section{Superluminality and instabilities}

The auxiliary field method that we have described could also shed light on the connection between the obstruction to UV-completion and the appearance of stable superluminal modes in the IR.  From the equation of motion for $\Phi$ in Eq. (\ref{Phiprop}) we have that, for perturbations about a background $\Phi_0$ and $\pi_0$,
\beq
\partial^2 (\delta \Phi) + c_3 \Lambda^2 \, \delta \Phi = - \frac{2 c_3 \epsilon}{\Lambda} \left[ (\partial_\mu \pi_0) \, \partial^\mu (\delta \pi) \right]~.
\label{PhiEOM}
\eeq
If \teq{\partial_\mu \pi_0 \neq 0}, then perturbations $\delta \pi$ lead to non-zero $\delta \Phi$, which is tachyonic for $c_3 < 0$.  We {\it conjecture} that superluminal perturbations in the IR are generally associated with instabilities near the cutoff scale.\footnote{This could be related to the instabilities near the cutoff scale that have been identified in ghost condensate models \cite{holes}.  As we have discussed, for certain backgrounds the ghost condensate also exhibits stable superluminal excitations in the IR, which could signal an obstruction to finding a high-energy completion for it (see \cite{low,donal}).}

Our approach may also rule out superluminal Goldstones in theories in which Lorentz invariance is spontaneously broken by a timelike vector VEV, \teq{ \left\langle S^\mu \right\rangle \neq 0}.  It should be pointed out, though, that the scattering of these Goldstones, being Lorentz non-invariant, is not adequately characterized by the kinematic variables $s$ and $t$.  The connection between our auxiliary field method and causality as encoded in the analytic structure of the S-matrix is not as transparent as in previous examples.

At the two-derivative level, the general effective action for such Goldstones can be written as
\beq
\La = c_1 \partial_{\alpha}S^{\beta} \partial^{\alpha} S_{\beta} +(c_2+c_3) \partial_{\mu} S^{\mu} \partial
_{\nu} S^{\nu} + c_4 S^{\mu} \partial_{\mu} S^{\alpha} S^{\nu}
\partial_{\nu} S_{\alpha} ~.
\label{EFT}
\eeq
If we normalize to \teq{S^2 =1} and work in the frame in which only $S^0$ is non-vanishing, we may write
\beq
S^{\mu}(x) \equiv \frac{1}{\sqrt{1-\vv \phi^2}} \left(1, \vv \phi \right)~,
\label{Smu}
\eeq
where $\vv \phi$ is as 3-vector whose components correspond to the Goldstones \cite{LV}.\footnote{Notice that the $S^\mu$'s are dimensionless while the $c_i$'s have mass dimension two.}  Classically, the linear Goldstone action is therefore
\beq
\La = \frac{1}{2} \sum_{i=1,2,3} \left[ \left(\partial_\mu \phi^i \right)^2 - \alpha \left( \partial_i \phi^i \right)^2 + \beta \left( \partial_0 \phi^i \right)^2 \right]~,
\label{eftgoldstones}
\eeq
where \teq{\alpha \equiv \left( c_2 + c_3 \right)/ c_1} and \teq{\beta \equiv c_4 / c_1}.  Absence of superluminal Goldstones requires \teq{\beta > 0} and \teq{\alpha < \beta}.  The action in Eq. (\ref{EFT}) is equivalent to
\beq
\La' = c_1 \left[ \partial_{\alpha}S^{\beta} \partial^{\alpha} S_{\beta} + 2 \alpha \Phi \left( \partial_\mu S^\mu \right) + 2 \beta A_\mu \left(S^\nu \partial_\nu S^\mu \right) - \alpha \Phi^2 - \beta A^2 \right]
+ \frac{1}{2} \left( \partial_\mu \Phi \right)^2 - \frac{1}{4} F_{\mu\nu}^2
\label{PhipropLV}
\eeq
for zero momentum of $\Phi_1$ and $A_\mu$.  Avoiding ghosts implies $c_1 < 0$.  Stability of Eq. (\ref{PhipropLV}) then requires that there be no superluminal Goldstones in Eq. (\ref{EFT}).\footnote{In fact, it also requires that the longitudinal mode, with \teq{v^2_{\sl{lgt}} = (1 + \alpha) / (1 + \beta)}, propagate more slowly than the two transverse modes, with \teq{v^2_{\sl{trv}} = 1/(1 + \beta)}.}  This observation might help to resolve the question of whether superluminal excitations should be forbidden or not in theories with spontaneous Lorentz violation \cite{eugene,jacobson}.  This issue is significant because the experimental constraint on spontaneous Lorentz violation coupled only to gravity is much tighter if superluminality of the Goldstones is forbidden \cite{nelson,moore}.

Analogously to what occurred in Eq. (\ref{PhiEOM}) for the scalar field, we see from the equation of motion for the action in Eq. (\ref{PhipropLV}) that stable superluminal Goldstones in Eq. (\ref{EFT}) are connected to excitations of tachyonic $\Phi$ or $A_\mu$.  We therefore conjecture that superluminal Goldstones are associated with instabilities that appear near the scale at which the spontaneously-broken Lorentz symmetry is restored.

\section{Conclusions}

We have described a very simple method for identifying a family of higher-dimensional operators in effective theories whose coefficient must be positive by causality:  We introduce auxiliary fields such that the original effective theory is reproduced when the auxiliary fields have zero momentum.  For operators of the form  \teq{{\cal O} = {\cal O}_1 \cdot {\cal O}_1}, where \teq{{\cal O}_1} contains enough derivatives, the positivity constraint on $\cal O$ from S-matrix analyticity is recast as a stability constraint on the sign of the mass-squared for the corresponding auxiliary field.

This procedure also identifies a family of operators for which causality alone should not impose positivity: those for which the theory with the auxiliary field can be stable and analytic regardless of sign of $\cal O$.  It is, of course, possible that there are other kinds of operators which must be positive by causality (or by another reason) but which our prescription does not detect.  For instance, some other positivity constraints which do not follow from our conjecture are obtained in \cite{lubos} from avoidance of ``Planck remnants'' (i.e., charged black holes that cannot decay quantum-mechanically).  For the operators which our conjecture does constrain, our results are consistent with \cite{lubos}.

We have also conjectured that what we have seen when applying our auxiliary field procedure is true in general: that stable superluminal modes in the IR of non-positive effective theories are connected to an instability that appears near the cutoff scale.  Finally, we have commented on what positivity constraints and causality imply for the ghost condensate, the chiral Lagrangian, and theories with spontaneous Lorentz violation.

\section{Acknowledgments}

We thank Allan Adams for a very useful and encouraging discussion, and for proposing the title.  We also thank Sean Carroll, Walter Goldberger, Michael Graesser, Andrey Katz, Eugene Lim, Delphine Perrodin, Surjeet Rajendran, and Mark Wise for valuable discussions and comments.  This work was supported in part by the U.S. Department of Energy under contracts DE-FG03-92ER40701 (A.J. and D.O.) and DE-FG02-05ER41360 (A.J.).


\end{document}